\documentclass[onecolumn,showpacs,preprintnumbers]{revtex4}
\usepackage{graphicx}
\usepackage{dcolumn}
\usepackage{bm}
\usepackage[english]{babel}
\begin{document}

\title{Localized states of an excess electron in ionic clusters.}
\author{K.V. Grigorishin}
\email{konst_dark@mail.ru}
\author{B.I. Lev}
\email{bohdan.lev@gmail.com} \affiliation{Boholyubov Institute for
Theoretical Physics of the Ukrainian National Academy of Sciences,
14-b Metrolohichna str. Kiev-03680, Ukraine.}
\date{\today}

\begin{abstract}
A theory for an electron affinity of ionic clusters is proposed
both in a quasiclassical approach and with quantization of a
polarization electric field in a nanoparticle. An interaction of
an electron with longitudinal optical phonons in an ionic cluster
is described. A critical size of the cluster regarding in
formation of an electron's autolocalized state, dependencies of an
energy and a radius of a polaron on a cluster's size are obtained
by a variational method. It has been found that a binding energy
of the electron in a cluster depends on a cluster's radius but the
radius of electron's autolocalization does not depend on cluster's
radius and equals to the polaron radius in a corresponding
infinity crystal, moreover the bound state of the electron in a
cluster is possible if the cluster's radius is more then the
polaron radius. A perturbation method for finding of the critical
parameters is proposed, where an dependence of the critical size
on a momentum of an electron in a conduction band is observed.

Keywords: cluster, polaron, autolocalization, phonon confinement.

%with an increase of the momentum the critical size decreases
%because of a strengthening of electron-phonon coupling with an
%increase of electron's energy.
\end{abstract}

\pacs{36.40.Wa, 63.20.kd, 63.22.Kn} \maketitle

\section{Introduction}\label{intr}

A bound state of an electron in a polar media is caused by a local
polarization of the medium inducted by the electron. The electron
is in a potential well in states with a discrete energy and its
field supports the polarization of a lattice. An interaction of
electrons with optical phonons resulting in an instability of an
electron's state in a conduction band is described by both a
quasiclassical approach and an electron-phonon Frohlich
hamiltonian. On basis of belief about a self-consistent state of
an electron and a polar medium (ammonium, water, protein globules)
mathematics models of an excess electron in a cluster have been
built and investigated \citep{stamp,lakhno,lakhno2}. In the models
it is supposed a solvated electron does not belong to a separate
molecule but collectively interacts with many atoms of the polar
medium. It is shown by analytical and numerical methods that a
critical size of molecular clusters consisting of polar molecules
(for example $(\texttt{H}_{2}\texttt{O})_{n}$ or
$(\texttt{NH}_{3})_{n}$) exists. In a cluster with a less size
than the critical size the autolocalized state (polaron state) can
not exist.

Since the polaron effect exists in an ionic crystal then we can
suppose the effect can occur in an ionic cluster too (for example
$\texttt{Na}_{14}\texttt{Cl}_{13}$) beginning with a some critical
size. In the papers \citep{land,honea1,rajag} electron
localization in alkali-halide clusters $M_{n}^{+}X_{n-1}^{-}$ and
$M_{n}^{+}X_{n-2}^{-}$ was numerically investigated. It was
supposed that the clusters contain F-center defects and electron
localized near an halide ion. Obviously, in this model an effect
of localization of an electron in a cluster takes place, but not
autolocalization because main contribution in a binding energy of
the excess electron is Coulomb-like interaction with F-center. The
critical size in this case is not observed. We will consider the
system neutral ionic cluster+excess electron where we have
interaction of the electron with an induced polarization only and
process of autolocalization.

The system electron+polarization can be described in a
quasiclassical approach. This means a motion of an electron is
quantized (is described by Schrodinger equation), however specific
polarization of a medium $\textbf{P}(\textbf{r})$ is considered as
a classical variable and described by electrostatic equations.
However an interaction of a electron with the polarization field
can be described in quantum approach.  Borders of a cluster
essentially influence upon quantization of optical oscillations -
the optical phonon confinement \citep{segi}, that is not
considered in the quasiclassical approach. Influence of a
geometric upon quantization of optical oscillations and an
interaction of an electron with them have been investigated in
papers \citep{licari,stros,cruz} for quantum wells, wires and
boxes. However, it should be noted that polaron states in the
heterostructures essentially depend on a confining potential. But
in a nanoparticle the polaron effect is stipulated by interaction
with an induced polarization only. A basic difference of optical
and acoustical modes in a cluster and heterostructure from the
modes in an infinity crystal is a restriction to possible
wavelengthes of the oscillation and presence of interfacial
phonons. These properties of phonons cause differences of a
polaron state in a nanoparticle from the polaron state in an
infinity crystal.

In a section \ref{quasi} we determine an energy of a polaron in a
nanoparticle and critical parameters of nonoparticles regarding to
formation of a polaron state by a variational method in a
quasiclassical approach. The variational method, unlike numerical
calculations, lets obtain analytical expressions for connection of
an autolocalized state's energy and a polaron radius with
parameters of a cluster (a size, an electron phonon coupling
constant, a frequency of lattice optical oscillations). In a
section \ref{strong} a field of deformations of an ionic cluster
is quantized taking into account the boundary condition on its
surface. The energy of a polaron state, the critical size of a
nanoparticle and the polaron radius are determined by a
variational method on the basis of an electron-phonon hamiltonian
for a spatially limited medium. Obtained results is compared with
result obtained by the numerical methods earlier. In a section
\ref{weak} a perturbation theory method was proposed in order to
find the critical parameters of a cluster.

\section{The quasiclassical approach.}\label{quasi}

In this section we will consider a quasiclassical description of a
electrons' interaction with a polarization displacement of ions
from their equilibrium positions. Let us suppose an electron is
localized in an ionic cluster of a spherical shape with a radius
$R$ and described by a wave function $\Psi(\textbf{r})$. A created
by electron mean Coulomb field
$\texttt{div}\textbf{D}(\textbf{r})=e|\Psi(\textbf{r})|^{2}$
induces a polarization of the cluster. In turn the polarization
electric field acts on the electron. A dipole moment of cluster's
unit volume $\textbf{P}(\textbf{r})$ is determined by a difference
of static and high-frequency polarizations:
\begin{equation}\label{1.1}
\textbf{P}(\textbf{r})=\textbf{P}_{0}(\textbf{r})-\textbf{P}_{\infty}(\textbf{r})
=\left(\frac{1}{\varepsilon_{\infty}}-\frac{1}{\varepsilon}\right)\textbf{D}\equiv\frac{1}{\widetilde{\varepsilon}}\textbf{D},
\end{equation}
where $\textbf{D}$ is an electric displacement, $\varepsilon$ and
$\varepsilon_{\infty}$ are static and high-frequency
permittivities accordingly. For a medium surrounding the cluster
(for example the vacuum) we have
$\varepsilon_{\infty}=\varepsilon$. Then
\begin{equation}\label{1.2}
\begin{array}{cc}
  \frac{1}{\widetilde{\varepsilon}}=\frac{\varepsilon-\varepsilon_{\infty}}{\varepsilon\varepsilon_{\infty}} & r<R \\
  \\
  \widetilde{\varepsilon}=\infty & r\geq R \\
\end{array}.
\end{equation}

A polarization of a cluster brings to appearance of a density of
polarization charge  $\rho=-\texttt{div}\textbf{P}$. Let
$\textbf{E}=-\nabla\varphi$ is an electric field of the
polarization, moreover
$\texttt{div}\textbf{E}=\rho/\varepsilon_{0}\Longrightarrow
\textbf{D}=-\varepsilon_{0}\widetilde{\varepsilon}\textbf{E}$,
where $\varepsilon_{0}$ is an electric constant. Then an energy of
a cluster's polarization is \cite{dav}:
\begin{equation}\label{1.3}
U_{\texttt{field}}=\frac{1}{2}\int\textbf{P}\frac{\textbf{D}}{\varepsilon_{0}}dV=
%\frac{1}{2}\int\frac{\textbf{D}^{2}}{\widetilde{\varepsilon}\varepsilon_{0}}dV=
\frac{1}{2}\varepsilon_{0}\int
\widetilde{\varepsilon}\textbf{E}^{2}dV.
\end{equation}
Hence an energy functional of a system electron+polarized cluster
has a form:
\begin{equation}\label{1.4}
    I(\Psi,\varphi)=\frac{\hbar^{2}}{2m}\int|\nabla\Psi|^{2}dV+e\int|\Psi|^{2}\varphi
    dV+\frac{1}{2}\varepsilon_{0}\int
\widetilde{\varepsilon}(\nabla\varphi)^{2}dV,
\end{equation}
where the first term is a kinetic energy of a localized electron
in the state $\Psi(\textbf{r})$, the second term is an interaction
energy of an electron with the electric field
$\varphi(\textbf{r})$ of the induced polarization. A localized
state of an electron in a cluster is energetically profitably if
$I<0$. Varying the functional (\ref{1.4}) with respect to $\Psi$
at a condition $\int|\Psi|^{2}dV=1$ we obtain an equation
describing movement of an electron with an energy $E$ in a
potential well $e\varphi$ created by the electron:
\begin{equation}\label{1.5}
    -\frac{h^{2}}{2m}\Delta\Psi+e\varphi\Psi=E\Psi.
\end{equation}
Varying the functional (\ref{1.4}) with respect to $\varphi$
taking into account (\ref{1.2}) we obtain an equation describing a
field of the polarization induced by the electron:
\begin{equation}\label{1.6}
\left\{
\begin{array}{cc}
  \Delta\varphi=\frac{e|\Psi|^{2}}{\varepsilon_{0}\widetilde{\varepsilon}} & r<R \\
  \\
  \Delta\varphi=0 & r\geq R \\
\end{array}
\right\}.
\end{equation}
Boundary conditions on a surface of the cluster (1) and the
environment (2) have a view:
\begin{equation}\label{1.7}
\varphi_{1}=\varphi_{2},\qquad
\widetilde{\varepsilon}_{1}\frac{\varphi_{1}}{\partial
n}=\widetilde{\varepsilon}_{2}\frac{\varphi_{2}}{\partial n},
\end{equation}
where $\textbf{n}$ is a normal to the cluster's surface. Since
$\widetilde{\varepsilon}_{2}=\infty$ and in view of
Eq.(\ref{1.6}), on a border of the cluster and out of the cluster
we can suppose
\begin{equation}\label{1.8}
    \varphi=0\qquad \texttt{for} \qquad r\geq R.
\end{equation}
The boundary condition (\ref{1.8}) $\varphi(R)=0$ is a main
difference of an electron's autolocalization in a cluster from a
polaron state in a infinity crystal where the electrical field of
a polarization is equal to zero at the infinity only:
$\varphi(\infty)=0$.

To obtain energies of polaron states and a critical size of a
cluster we will use a variational method because it lets find
analytical expressions. Let the cluster is characterized by values
of static $\varepsilon$ and high-frequency $\varepsilon_{\infty}$
permittivities and effective mass $m$ of an electron in a
crystal's conduction band (for $\texttt{NaCl}$ $m=2.78m_{0}$,
$m_{0}$ is a electron's mass in vacuum). Obviously these values
have sense for clusters with sizes which are much more than
interatomic distances ($\sim 2\div 3\texttt{A}$) only. However a
proposed below method can be extrapolated to small clusters. With
help of the equations (\ref{1.6}) and the condition
$\varphi(\infty)=0$ the functional (\ref{1.4}) can be simplified:
\begin{equation}\label{1.9}
    I(\Psi,\varphi)=\frac{\hbar^{2}}{2m}\int|\nabla\Psi|^{2}dV-\frac{1}{2}\varepsilon_{0}\int
\widetilde{\varepsilon}(\nabla\varphi)^{2}dV.
\end{equation}
For a region out of the cluster $r\geq R$ we suppose
$\widetilde{\varepsilon}\varphi=0$. A wave function of a system's
(electron+polarized cluster) ground state can be selected in a
form \cite{dav}:
\begin{equation}\label{1.10}
    \Psi=\frac{1+r/r_{0}}{\sqrt{7\pi
    r_{0}^{3}}}\exp\left(-\frac{r}{r_{0}}\right),
\end{equation}
where the radius $r_{0}$ is a variational parameter - a polaron
radius. As far as the cluster is spherical, then the equation
(\ref{1.6}) can be rewritten in a view:
\begin{equation}\label{1.12}
\frac{1}{r^{2}}\frac{\partial}{\partial
r}\left(r^{2}\frac{\partial\varphi}{\partial
r}\right)=\frac{e|\Psi|^{2}}{\varepsilon_{0}\widetilde{\varepsilon}}.
\end{equation}
The first integral and a solution of this equation are functions:
\begin{eqnarray}
    \frac{\partial\varphi}{\partial r}&=&\frac{e}{\varepsilon_{0}\widetilde{\varepsilon}}
    \left(\frac{1}{r^{2}}\left(\int|\Psi|^{2}r^{2}dr+C_{1}\right)\right)\label{1.13}\\
    \varphi&=&\frac{e}{\varepsilon_{0}\widetilde{\varepsilon}}
    \left(\int\frac{1}{r^{2}}\left(\int|\Psi|^{2}r^{2}dr+C_{1}\right)dr+C_{2}\right),\label{1.14}
\end{eqnarray}
where the constant $C_{1}$ must be selected for
$\varphi(0)\neq\infty$, the constant $C_{2}$ must be found from
the boundary condition (\ref{1.8}). Substituting the wave function
(\ref{1.10}) in Eqs.(\ref{1.13},\ref{1.14}) we have:
\begin{eqnarray}
    \frac{\partial\varphi}{\partial r}&=&\frac{e}{\varepsilon_{0}\widetilde{\varepsilon}}\frac{1}{4\pi
    r^{2}}\left[1-\exp\left(-2\frac{r}{r_{0}}\right)\left(1+2\frac{r}{r_{0}}+2\frac{r^{2}}{r_{0}^{2}}
    +\frac{8}{7}\frac{r^{3}}{r_{0}^{3}}+\frac{2}{7}\frac{r^{4}}{r_{0}^{4}}\right)\right],
    \qquad \texttt{for} \qquad r\leq R
    \label{1.15}\\
    \varphi&=&\frac{e}{\varepsilon_{0}\widetilde{\varepsilon}}\frac{14}{56\pi
    r}\left[\exp\left(-2\frac{r}{r_{0}}\right)\left(1+\frac{19}{14}\frac{r}{r_{0}}+\frac{10}{14}\frac{r^{2}}{r_{0}^{2}}
    +\frac{2}{14}\frac{r^{3}}{r_{0}^{3}}\right)-1\right]\nonumber\\
    &-&\frac{e}{\varepsilon_{0}\widetilde{\varepsilon}}\frac{14}{56\pi
    R}\left[\exp\left(-2\frac{R}{r_{0}}\right)\left(1+\frac{19}{14}\frac{R}{r_{0}}+\frac{10}{14}\frac{R^{2}}{r_{0}^{2}}
    +\frac{2}{14}\frac{R^{3}}{r_{0}^{3}}\right)-1\right],\qquad \texttt{for} \qquad r\leq
    R\label{1.16}\\
    &&\frac{\partial\varphi}{\partial r}=0, \qquad \varphi=0, \qquad \texttt{for} \qquad
    r>R.\label{1.17}\
\end{eqnarray}
Substituting $\Psi$ (\ref{1.10}) and
$\frac{\partial\varphi}{\partial r}$ (\ref{1.15},\ref{1.17}) in
the functional $I(\Psi,\varphi)$ (\ref{1.9}) we obtain an energy
of a polarized cluster with a localized electron as a function of
a polaron radius:
\begin{eqnarray}\label{1.18}
    I(r_{0})&=&\frac{3\hbar^{2}}{14mr_{0}^{2}}-\frac{5373}{100352}\frac{e^{2}}{\varepsilon_{0}\widetilde{\varepsilon}\pi r_{0}}
    \left[1+\frac{\exp\left(-2\frac{R}{r_{0}}\right)}{5373r_{0}^{6}R}
    \left(34048r_{0}^{6}R-25088r_{0}^{7}+17920r_{0}^{5}R^{2}+3584r_{0}^{4}R^{3}\right)\right]\nonumber\\
    &+&\frac{e^{2}}{\varepsilon_{0}\widetilde{\varepsilon}\pi r_{0}}\frac{\exp\left(-4\frac{R}{r_{0}}\right)}{100352r_{0}^{6}R}
    (12544r_{0}^{7}+39421r_{0}^{6}R+57322r_{0}^{5}R^{2}+50152r_{0}^{4}R^{3}+28640r_{0}^{3}R^{4}+10720r_{0}^{2}R^{5}\nonumber\\
    &+&2432r_{0}R^{6}+256R^{7})+\frac{12544}{100352}\frac{e^{2}}{\varepsilon_{0}\widetilde{\varepsilon}\pi
    r_{0}}\frac{r_{0}}{R}.
\end{eqnarray}
The following terms give a main contribution in the energy
(\ref{1.18}) at $r_{0}R>0.5$:
\begin{equation}\label{1.19}
    I(r_{0})\approx\frac{3\hbar^{2}}{14mr_{0}^{2}}-\frac{5373}{100352}\frac{e^{2}}{\varepsilon_{0}\widetilde{\varepsilon}\pi r_{0}}
    \left[1-\frac{12544}{5373}\frac{r_{0}}{R}\right]=I_{\infty}+\frac{12544}{100352}\frac{e^{2}}{\varepsilon_{0}\widetilde{\varepsilon}\pi
    R},
\end{equation}
where $I_{\infty}$ is an energy of an infinity crystal with a
localized electron. An autolocalized state of an electron in a
cluster (or in a crystal) is energetically profitable if $I<0$.
From Eqs.(\ref{1.18},\ref{1.19}) we can see a spatial limitation
of a crystal increases the energy of the polaron state:
$I_{\infty}<I$. From Eq.(\ref{1.19}) we can see the polaron's
radius $r_{0}$ does not depend on the cluster's radius $R$:
\begin{equation}\label{1.20}
    \frac{\partial I}{\partial r_{0}}=0\Longrightarrow r_{0}=\frac{3}{7}\frac{100352}{5373}
    \frac{\hbar^{2}\varepsilon_{0}\widetilde{\varepsilon}\pi}{me^{2}}.
\end{equation}

\begin{figure}[h]
\includegraphics[width=8.6cm]{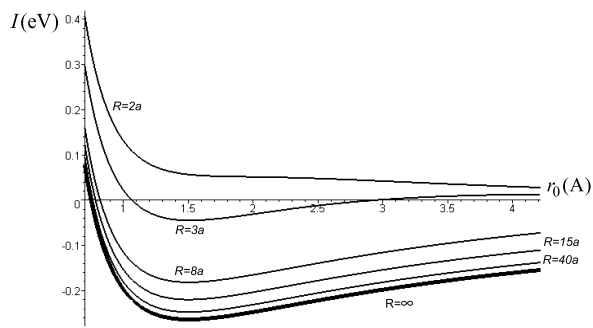}
\caption{The dependencies of an energy of a system
electron+cluster (\ref{1.18}) on a polaron radius $I=I(r_{0})$ for
an ionic cluster consisting of molecules $\texttt{NaCL}$. The
curves have been built for cluster's radii: $2a,3a,8a,15a,40a$,
where $a=2.81\texttt{A}$ is an interatomic distance. The bold line
is the energy of an infinity crystal with an electron.}
\label{fig1}
\end{figure}

The function $I(r_{0})$ (\ref{1.18}) for different values of the
cluster's radii R is shown in Fig.(\ref{fig1}). We can see, that a
bound state of an electron in a cluster consisting of molecules
$\texttt{NaCL}$ can exist beginning from $R\approx 3a$, where
$a=2.81\texttt{A}$ is an interatomic distance. With increase a
cluster's size a bound energy increases and aspires to a polaron
energy in an infinity crystal
$I_{\infty}\approx-0.265\texttt{eV}$. A radius of electron's
autolocalization $r_{0}$ in a cluster minimizes the energy $I$.
From Fig.(\ref{fig1}) we can see $r_{0}$ does not depend on a
cluster's radius almost and for sodium chloride the radius is
equal to $r_{0}=1.5\texttt{A}$. The dependencies shown in
Fig.(\ref{fig1}) is well approximated by the more simple
dependence (\ref{1.19}).

\begin{figure}[h]
\includegraphics[width=8.6cm]{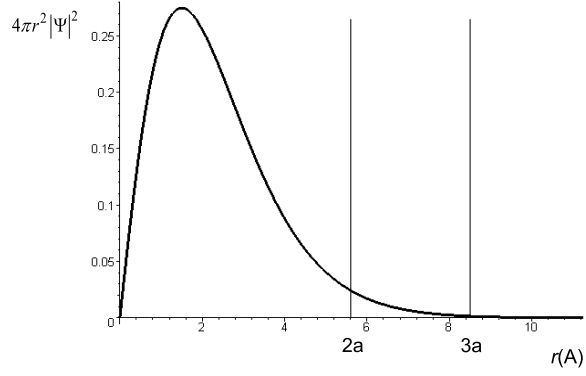}
\caption{The probability distribution of being of an electron in a
cluster. The vertical lines means radii of clusters for which an
energy of an electron's autolocalized state was calculated in
Fig.(\ref{fig1}). The radius $2a$ is subcritical (the bound state
is absent), the radius $3a$ is some bigger than the critical
radius - the bound state appears.} \label{fig2}
\end{figure}

The equations (\ref{1.5},\ref{1.6}) were solved numerically in a
paper \cite{lakhno} for clusters of ammonia and water. It was
shown that critical size of the clusters consisting of polar
medium exists. However analytical method discovers the next
pattern. The polaron radius in a cluster does not depend on a
radius of the cluster. The bound state of an electron in the
cluster exists till a polaron radius is less than the cluster's
radius. That is the autolocalized electron is hidden in depth of
the cluster always and has a negligibly low probability to be on
the outside. If the cluster is enough small so that the electron
can be out of the cluster, the bound state disappears.
Above-mentioned particulars are illustrated in Fig.(\ref{fig2}).
Similar result ware obtained in \cite{bar0}, but another
functional $I(\Psi,\varphi)$ was considered and ground state
energy of an electron in a cluster and the autolocalization radius
were expressed by static dielectric constant only. Hence in this
model autolocalisation effect is not polaron effect unlike our
model.

\begin{figure}[h]
\includegraphics[width=8.6cm]{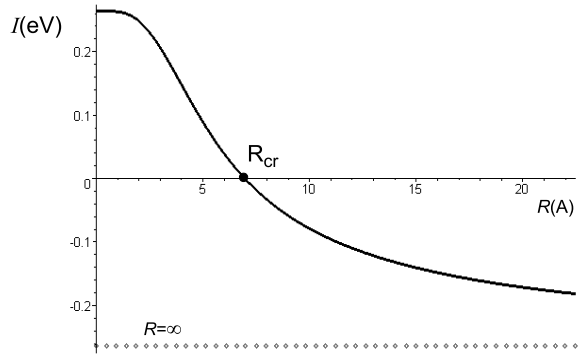}
\caption{The energy of a system electron+cluster (consisting of
$\texttt{NaCL}$) as a function of a cluster's radius $I(R)$ (the
solid line). The doted line is an energy of a polaron in an
infinite crystal $R=\infty$. $R_{\texttt{cr}}$ is the critical
radius of a cluster.} \label{fig3}
\end{figure}

Since a polaron radius $r_{0}$ is independent on a cluster's
radius $R$, then we can obtain an energy of a system
electron+cluster as a function of the cluster's radius $I(R)$
assuming in the formula (\ref{1.18}) the polaron radius is equal
to the optimal value (\ref{1.20}). The result is shown in
Fig.(\ref{fig3}), where we can see the critical radius of a
cluster (consisting of $\texttt{NaCL}$) determined by the equation
$I(R_{\texttt{cr}})=0$ is equal to $R_{\texttt{cr}}=7.0\texttt{A}$
($2\div 3$ interatomic distances). With  increase of the cluster's
radius the energy asymptotically aspires to its value in an
infinite crystal $I(R\rightarrow\infty)\rightarrow I_{\infty}$.

\begin{figure}[h]
\includegraphics[width=8.6cm]{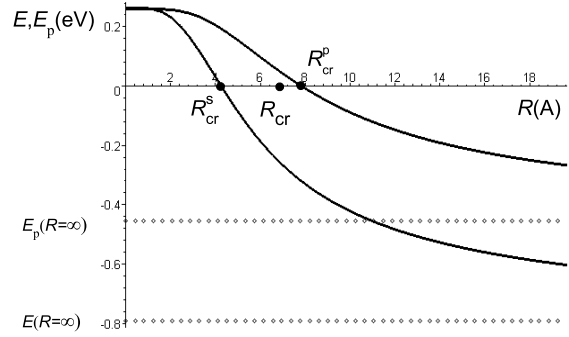}
\caption{The energies of a electron's ground state $1\texttt{s}$
and an electron's $\texttt{p}$-state in the potential well
$e\varphi$ as functions of a cluster's radius $E(R)$ and
$E_{p}(R)$ (solid lines). The dotted lines show the energies of an
electron in an infinity crystal $R=\infty$. $R_{\texttt{cr}}$ is a
critical radius when the polaron state is energetically profitable
(see Fig.\ref{fig3}), $R_{\texttt{cr}}^{\texttt{s}}$ and
$R_{\texttt{cr}}^{p}$ are a critical radii when $1s$-state and
$p$-state of an electron in the potential well
(\ref{1.16},\ref{1.17}) is energetically profitable accordingly.}
\label{fig4}
\end{figure}

A ground state of an electron in the potential well $e\varphi$ is
$1\texttt{s}$. The energy $I(\Psi,\varphi)$ (\ref{1.4},\ref{1.9})
is a sum of a ground state energy of an electron in the potential
well and an energy of a cluster's polarization
$1/2\varepsilon_{0}\int
\widetilde{\varepsilon}(\nabla\varphi)^{2}dV>0$. Then the ground
state energy of an electron in a cluster is
\begin{equation}\label{1.21}
    E(R)=\frac{\hbar^{2}}{2m}\int|\nabla\Psi|^{2}dV-\varepsilon_{0}\int
\widetilde{\varepsilon}(\nabla\varphi)^{2}dV,
\end{equation}
moreover the wave function $\Psi$ and the field $\varphi$ is taken
in the forms (\ref{1.10}) and (\ref{1.15}-\ref{1.17}) accordingly.
The radius $r_{0}$ is equal to its optimal value (\ref{1.20}). The
energy $E(r)$ is a necessary energy to quickly remove an electron
from the localized state to a free state (for example by
absorption of a photon with an energy $\hbar\omega>|E|$). The
graph $E(R)$ is shown in Fig.(\ref{fig4}), where we can see that
$1s$-state is energetically profitable beginning from a radius of
the cluster $R_{\texttt{cr}}^{s}\approx
4\texttt{A}<R_{\texttt{cr}}=7\texttt{A}$ (substance
$\texttt{NaCl}$). That is in the interval
$R_{\texttt{cr}}^{s}<R<R_{\texttt{cr}}$ a metastable autolocalized
state of an electron in an ionic cluster is possible.

\begin{figure}[h]
\includegraphics[width=8.6cm]{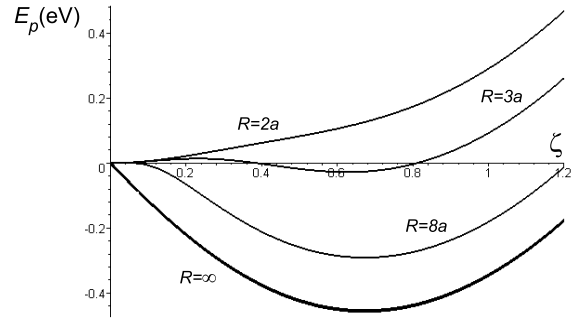}
\caption{The dependence of electron's energy in $p$-state in a
cluster on a parameter $\zeta$ for an ionic cluster consisting of
$\texttt{NaCL}$. The curves have been built for radii of clusters:
$2a,3a,8a$, where $a=2.81\texttt{A}$ is an interatomic distance.
The bold line is electron's energy in $p$-state in an infinite
crystal.} \label{fig5}
\end{figure}

If the potential well $e\varphi$ (\ref{1.16}) is sufficiently deep
then other discrete energy levels are possible. At increasing of
the cluster's radius $p$-state can appear. $s-p$ photo-transitions
occur without changes of ions' positions - Franck–Condon
principle. That is the electrical field $\varphi$ of a
polarization and a polaron radius $r_{0}$ are the same as the
values in $s$-state: (\ref{1.16}) and (\ref{1.20}) accordingly,
however a wave function of an electron can be chosen in a form:
\begin{equation}\label{1.22}
    \Psi=\left(\frac{\zeta}{r_{0}}\right)^{3/2}\frac{r\zeta}{\pi r_{0}}\exp\left(-\frac{\zeta r}{r_{0}}\right)\cos\theta,
\end{equation}
where $\zeta$ is a variational parameter, $\theta$ is a polar
angle. Then an energy of an electron in $p$-state is:
\begin{equation}\label{1.23}
    E_{p}=\frac{\hbar^{2}}{2m_{\ast}}\int_{0}^{\infty}2\pi
    r^{2}dr\int_{0}^{\pi}\sin\theta
    d\theta\left|\nabla\Psi_{p}(r,\theta)\right|^{2}+e\int_{0}^{\infty}2\pi
    r^{2}dr\int_{0}^{\pi}\sin\theta
    d\theta\left|\Psi_{p}(r,\theta)\right|^{2}\varphi(r).
\end{equation}
The expression (\ref{1.23}) is a function of a variation parameter
$\zeta$. The graphics of a dependence $E_{p}$ on $\zeta$ for
clusters' radii $R=2a,3a,8a,\infty$ are shown in Fig.(\ref{fig5}),
where we can see an optimal value (it minimizes $E_{p}$) of the
parameter $\zeta$ weakly depends on a cluster's radius and it can
be supposed $\zeta=0.65$. Then we can easily find a dependence of
$\texttt{p}$-state energy on a cluster's radius $E_{p}(R)$ -
Fig.(\ref{fig4}), where we can see $p$-state is energetically
profitable beginning from the radius of a cluster
$R_{\texttt{cr}}^{p}\approx
8\texttt{A}>R_{\texttt{cr}}>R_{\texttt{cr}}^{s}$. With increasing
of a cluster's radius the energy $E_{p}(R)$ asymptotically aspires
to its value in an infinity crystal. In a bigger clusters $d$, $f$
and so on states are possible.

It necessary to notice that all found critical sizes for the
cluster have values of the order of several interatomic distances
$a$. This means the continuous medium model is not good for
description of a such nanoparticle. Moreover an effective mass $m$
in a conduction band figures in the model. Hence the model is
correct for enough big clusters $R\gg a$. Sodium chloride is a
strongly polarizable substance. For less polarizable substances
the model can be used with a better accuracy.

An essential difference of a polaron state of an electron in a
cluster from the polaron state in an infinity crystal is
localization of an electron in the cluster's center in result of
the boundary condition (\ref{1.8}). On the contrary in an infinity
crystal all points are equivalent (in a continuous representation
of a substance) therefore the polaron can move. In this case the
polaron is characterized by an effective mass $M\gg m$.

\section{The quantum approach. A strong coupling theory of an electron in a cluster.}\label{strong}

An interaction of a electron with the polarization field can be
described in quantum approach.  Borders of a cluster essentially
influence upon quantization of optical oscillations that is not
considered in the quasiclassical approach. A basic difference of
optical and acoustical modes in a cluster from the modes in an
infinity crystal is a restriction to possible wavelengthes of the
oscillation. In a such situation it is necessary to quantize a
field of a cluster's polarization $\varphi$. In a harmonic
approximation a hamiltonian of a system electron+cluster has a
form:
\begin{equation}\label{2.1}
    H=-\frac{\hbar^{2}}{2m}\Delta+
    \frac{1}{2}\varepsilon_{0}\widetilde{\varepsilon}\int\left[\frac{1}{\omega^{2}}(\nabla\dot{\varphi})^{2}+(\nabla\varphi)^{2}\right]dV+
    e\varphi\equiv\widehat{T}+\widehat{U}_{f}+\widehat{V}_{int},
\end{equation}
where the first term $\widehat{T}$ is an operator of a kinetic
energy of an electron, $\widehat{U}_{f}$ is an operator of a
deformation energy of a crystal (cluster), $\widehat{V}_{int}$ is
an operator of an interaction of the electron with the deformation
field. The procedure of quantization of the deformation field lies
in the fact that the electric field $\varphi$ must be expressed in
creation $b^{\dagger}$ and annihilation $b$ operators of optical
phonons so that
\begin{equation}\label{2.2}
    \widehat{U}_{f}=
    \frac{1}{2}\varepsilon_{0}\widetilde{\varepsilon}\int\left[\frac{1}{\omega^{2}}(\nabla\dot{\varphi})^{2}+(\nabla\varphi)^{2}\right]dV=
    \hbar\omega\sum_{\textbf{q}}\left(b_{\textbf{q}}^{\dagger}b_{\textbf{q}}+\frac{1}{2}\right),
\end{equation}
where $\textbf{q}$ is a wave vector of a phonon. For simplicity we
will suppose optical oscillations is dispersionless
$\omega=\texttt{const}$. The field $\varphi$ is proportional to a
displacement of atoms $\xi$. The displacement and the field must
satisfy periodic boundary conditions:
\begin{equation}\label{2.3}
    \xi_{\textbf{n}}=\xi_{\textbf{n}+N2\textbf{a}},\qquad
    \varphi_{\textbf{n}}=\varphi_{\textbf{n}+N2\textbf{a}},
\end{equation}
where $2a$ is a lattice constant of a two-component ionic crystal.
Then interaction of an electron with the deformation field
(electron-phonon interaction) is written as follows:
\begin{equation}\label{2.4}
    \widehat{V}_{int}=e\widehat{\varphi}=e\sqrt{\frac{\hbar\omega}{2\varepsilon_{0}\widetilde{\varepsilon}V}}
    \sum_{\textbf{q}}\frac{e^{i\textbf{qr}}}{q}\left(b_{\textbf{q}}+b_{-\textbf{q}}^{\dagger}\right),
\end{equation}
where $\textbf{q}$ ia taken from the first Brillouin zone
$-\pi/2a<q_{x},q_{y},q_{z}<\pi/2a$. The creation and annihilation
operators are written in sense $b_{\textbf{q}}\rightarrow
b_{\textbf{q}}e^{i\omega t}$, $b_{\textbf{q}}^{\dag}\rightarrow
b_{\textbf{q}}^{\dag}e^{-i\omega t}$. The cyclic boundary
conditions are ensured by periodic multipliers $e^{i\textbf{qr}}$.
For an infinity crystal we can neglect discreteness of a
substance and we can assume a period of a lattice is infinitely
small, hence $-\infty<q_{x},q_{y},q_{z}<\infty$.

Due ionic type connection between atoms a shape of ionic cluster
is close to cubic (for example unlike a ball shape of metallic
cluster). Computer simulations and experimental evidences
\cite{land,honea1,rajag} speak about alkali halide clusters
$M_{n}^{+}X_{m}^{-}, n\approx m\gtrsim 14$ form stable cuboid
structures and have a cubic microlattice. Clusters as and
nucleuses can have magic numbers of atoms - such numbers of atoms
when all shells are filled. Magic clusters are most stable and
magic ionic clusters have cuboid shape.  For large cluster $R\gg
a$ a shape can any ($a$ is an interatomic distance). Obviously for
large clusters $R\gg r_{0}$ the shape have not principal
importance because an excess electron has low probability to be
near a surface of the cluster. For convenience we will consider
the large cluster as cubic in this case too.

Let we have an ionic cluster (for example consisting of
$\texttt{NaCL}$) of a cubic form with sizes
$-\frac{L}{2}<x,y,z<\frac{L}{2}$. A main difference of a cluster
from an infinity crystal is the condition on a cluster's boundary
(\ref{1.8}) instead of the cyclic boundary condition (\ref{2.3}),
that causes the phonon confinement - Fig.\ref{fig7}. That is for
the cube the boundary conditions have a view:
\begin{equation}\label{2.5}
    \varphi\left(x=\pm\frac{L}{2}\right)=\varphi\left(y=\pm\frac{L}{2}\right)=\varphi\left(z=\pm\frac{L}{2}\right)=0.
\end{equation}
The field $\varphi$ (\ref{2.4}) does not satisfy the conditions
(\ref{2.5}). In the article \cite{licari} a method of construction
of a hamiltonian for an electron-optical phonon interaction in a
dielectric slab (with boundary conditions
$\varphi\left(z=\pm\frac{L}{2}\right)=0$) was proposed. But for a
quantum box (cluster) the method brings to a very cumbersome
expression. We propose the following method. The boundary
condition is satisfied by the operator
\begin{eqnarray}\label{2.6}
    \widehat{\varphi}&=&\frac{A}{\sqrt{L^{3}}}
    \sum_{q_{x}}\sum_{q_{y}}\sum_{q_{z}}\frac{1}{q}\left(b_{\textbf{q}}+b_{-\textbf{q}}^{\dagger}\right)\nonumber\\
    &\times&
    \left[%
\begin{array}{cc}
  \cos(q_{x}x), & \texttt{odd}\qquad n_{x} \\
  \sin(q_{x}x), & \texttt{even}\qquad n_{x} \\
\end{array}%
\right]
    \left[%
\begin{array}{cc}
  \cos(q_{y}y), & \texttt{odd}\qquad n_{y} \\
  \sin(q_{y}y), & \texttt{even}\qquad n_{y} \\
\end{array}%
\right]
    \left[%
\begin{array}{cc}
  \cos(q_{z}z), & \texttt{odd}\qquad n_{z} \\
  \sin(q_{z}z), & \texttt{even}\qquad n_{z} \\
\end{array}%
\right],
\end{eqnarray}
where projections of a wave vector $\textbf{q}$ of optical phonon
are $q_{x,y,z}\frac{L}{2}=n_{x,y,z}\frac{\pi}{2}$ (such that
$\cos\left(q_{x,y,z}\frac{L}{2}\right)=0,\qquad\sin\left(q_{x,y,z}\frac{L}{2}\right)=0$).
The unknown constant $A$ is found from the condition of secondary
quantization of the deformation field (\ref{2.2}). From where we
obtain
$A=\sqrt{\frac{\hbar\omega}{2\varepsilon_{0}\widetilde{\varepsilon}}}$.
Energy of an electron is minimal when it is situated in the center
of a cluster and a probability amplitude is symmetrical about the
center $\Psi(\textbf{r})=\Psi(-\textbf{r})$. Then the summands
with $\sin$ may be omitted because they do not give any
contribution in an electron-phonon interaction energy (and hence
in a polarization - the field $\varphi$):
$\int_{-L/2}^{L/2}\Psi^{2}(x)\sin(q_{x}x)=0$. This is equivalent
to boundary conditions
\begin{equation}\label{2.5a}
    \frac{\partial\varphi}{\partial x}(x=0)=\frac{\partial\varphi}{\partial y}(y=0)=\frac{\partial\varphi}{\partial z}(z=0)=0.
\end{equation}
The field $\varphi$ must satisfy the boundary conditions
(\ref{2.5},\ref{2.5a}) on a cluster's surface. Whole number $n$ of
half-waves of phonons must be placed in a cube edge $L=na$.
Moreover a half of phonon's wavelength must not be bigger than
length of a cube edge and smaller than interatomic distance $a$
(between a cation and an anion on a cube edge)
$a<\lambda/2=\frac{\pi}{q}<L$. Hence the possible projections of
wave vector $\textbf{q}=(q_{x},q_{y},q_{z})$ of an optical phonon
must take on values:
\begin{eqnarray}\label{2.7}
\left(%
\begin{array}{c}
  q_{x}\frac{L}{2}=n_{x}\frac{\pi}{2} \\
  \\
  q_{y}\frac{L}{2}=n_{y}\frac{\pi}{2} \\
  \\
  q_{z}\frac{L}{2}=n_{z}\frac{\pi}{2} \\
\end{array}%
\right),\qquad n_{x},n_{y},n_{z}=\pm 1,\pm3,\pm 5,\ldots,\left(%
\begin{array}{cc}
  \pm n, & \texttt{odd}\qquad n \\
  \pm (n-1), & \texttt{even}\qquad n \\
\end{array}%
\right),\qquad L=na
\end{eqnarray}

Then a hamiltonian of a system electron+cluster is written in a
view:
\begin{equation}\label{2.8}
    H=-\frac{\hbar^{2}}{2m}\Delta+
    \hbar\omega\sum_{\textbf{q}}\left(b_{\textbf{q}}^{\dagger}b_{\textbf{q}}+\frac{1}{2}\right)+
    \frac{M_{0}}{\sqrt{L^{3}}}
    \sum_{\textbf{q}}\frac{\cos(q_{x}x)\cos(q_{y}y)\cos(q_{z}z)}{q}\left(b_{\textbf{q}}+b_{-\textbf{q}}^{\dagger}\right),
\end{equation}
where
\begin{equation}\label{2.9}
    M_{0}^{2}=e^{2}\frac{\hbar\omega}{2\varepsilon_{0}\widetilde{\varepsilon}}\equiv\frac{4\pi\alpha\hbar(\hbar\omega)^{3/2}}{\sqrt{2m}}
    ,\qquad \alpha=\frac{e^{2}}{4\pi\varepsilon_{0}\widetilde{\varepsilon}\hbar}\left(\frac{m}{2\hbar\omega}\right)^{1/2}.
\end{equation}
Here $\alpha$ plays a role in a electron-phonon coupling constant.
We suppose optical phonons are dispersionless and $\alpha$ does
not depend on sizes of a cluster. Experimental data say about an
oscillatory spectrum $\omega(q)$ for clusters
$(\texttt{NaCL})_{N}$ quickly approaches with an oscillatory
spectrum of an infinity crystal at increase of $N$. So $N=4$ is
enough for having characteristics of an infinity crystal
\cite{welch}. Since an autolocalized electron is hidden in a depth
of the cluster always and has a negligibly low probability to be
on the outside (it was shown in a previous section) then we can
neglect of an interaction with interface phonons.

Let us find specific polarization operator $\textbf{P}$ using a set
of equations:
\begin{equation}\label{2.10}
    \left\{%
\begin{array}{c}
  \Delta\varphi=-\frac{\rho}{\varepsilon_{0}} \\
  \\
  \rho=-\texttt{div}\textbf{P} \\
\end{array}%
\right\}\Longrightarrow\textbf{P}=\varepsilon_{0}\nabla\varphi,
\end{equation}
where $\rho$ is a density of polarization charges. Then
\begin{eqnarray}\label{2.11}
    \textbf{P}=&&\sqrt{\frac{\hbar\omega}{2\varepsilon_{0}\widetilde{\varepsilon}L^{3}}}
    \sum_{\textbf{q}}\frac{1}{q}\left(b_{\textbf{q}}+b_{-\textbf{q}}^{\dag}\right)\nonumber\\
    &&\times\left[\textbf{i}q_{x}\sin(q_{x}x)\cos(q_{y}y)\cos(q_{z}z)
    +\textbf{j}q_{y}\cos(q_{x}x)\sin(q_{y}y)\cos(q_{z}z)
    +\textbf{k}q_{z}\cos(q_{x}x)\cos(q_{y}y)\sin(q_{z}z)\right].
\end{eqnarray}
In turn, the specific polarization vector is proportional to
displacements of atoms of a crystal (cluster):
$\textbf{P}=\frac{Ne\xi}{V\varepsilon_{\infty}}$. The fields
$\varphi$ and $\xi$ can be qualitatively illustrated by the
example of an atom chain. Then in 1D system we have
$\varphi\propto\cos(qx)$, $\xi\propto q\sin(qx)$. These
configurations of the fields are shown schematically in
Fig.(\ref{fig7}). We can see in a chain consisting of four atoms
(two anions and two cations) the oscillations are possible with
two wave vectors $q=\pi/3a$ and $q=\pi/a$ only in the course of
the boundary conditions (\ref{2.5},\ref{2.5a}).
\begin{figure}[h]
\includegraphics[width=8.6cm]{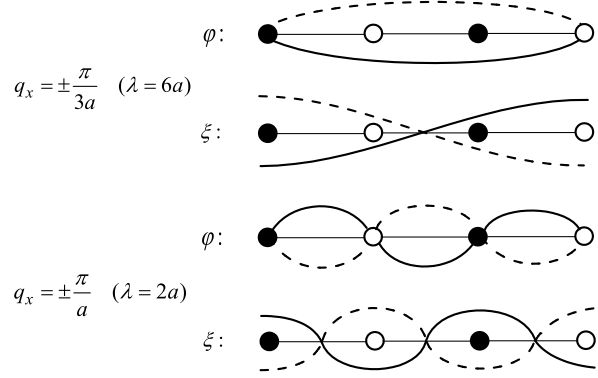}
\caption{The schematic illustration of possible oscillations
(in the course of the boundary conditions (\ref{2.5},\ref{2.5a}))
of an electric polarization field $\varphi$ and an atom displacement $\xi$
for a chain consisting of two cations and two anions
%(the black and white circulus)
.} \label{fig6}
\end{figure}

It is convenient to transit from the creation and annihilation
operators to new conjugate operators:
\begin{equation}\label{2.12}
    Q_{\textbf{q}}=\frac{1}{\sqrt{2}}\left(b_{\textbf{q}}+b_{-\textbf{q}}^{\dag}\right),\qquad
    P_{\textbf{q}}=-\frac{i}{\sqrt{2}}\left(b_{\textbf{q}}-b_{-\textbf{q}}^{\dag}\right).
\end{equation}
Then the hamiltonian is written in a view:
\begin{eqnarray}\label{2.13}
\widehat{H}=&&-\frac{\hbar^{2}}{2m_{\ast}}\Delta+\frac{\hbar\omega}{2}\sum_{\textbf{q}}\left(|Q_{\textbf{q}}|^{2}+|P_{\textbf{q}}|^{2}\right)\nonumber\\
&&+M_{0}\sqrt{\frac{2}{L^{3}}}\sum_{\textbf{q}}\frac{\cos(q_{x}x)\cos(q_{y}y)\cos(q_{z}z)}{q}Q_{\textbf{q}}.
\end{eqnarray}
The system's wave function $\Phi(\textbf{r},Q_{\textbf{q}})$ has
to contain coordinates of an electron $\textbf{r}$ and ions'
displacement $Q_{\textbf{q}}$. Usually they suppose the wave
function is product of two parts depending only on electron coordinates
and phonon coordinates accordingly:
\begin{eqnarray}
  &\Phi(\textbf{r},Q_{\textbf{q}}) = \Psi(\textbf{r})\phi(Q_{\textbf{q}}+\delta Q_{\textbf{q}}) \label{2.14}\\
  &\Psi(\textbf{r}) = \left(\frac{1}{\pi
  r_{0}}\right)^{3/2}\exp\left(-\frac{r^{2}}{2r_{0}^{2}}\right)\label{2.15},
\end{eqnarray}
where the electron wave function is chosen in Gauss view to
describe a localized electron in a cluster, $r_{0}$ is a polaron
radius playing a role of a variational parameter. The phonon wave
function $\phi$ is a wave function of a harmonic oscillator
centered on the equilibrium displacement $-\delta Q_{\textbf{q}}$
(an electron deforms a cluster and ions displace in new centers of
equilibrium), which must be determined. Let us average the
hamiltonian (\ref{2.13}) over electron's coordinates:
$H(Q_{\textbf{q}})=\int
\Psi\dag(\textbf{r})\widehat{H}\Psi(\textbf{r})d^{3}r$:
\begin{eqnarray}\label{2.16}
H(Q_{\textbf{q}})=\frac{\hbar\omega}{2}\sum_{\textbf{q}}\left(|Q_{\textbf{q}}|^{2}+|P_{\textbf{q}}|^{2}\right)
+\frac{3\hbar^{2}}{4m_{\ast}r_{0}^{2}}+\sum_{\textbf{q}}L_{\textbf{q}}Q_{\textbf{q}},
\end{eqnarray}
where
\begin{equation}\label{2.17}
    L_{\textbf{q}}=M_{0}\sqrt{\frac{2}{L^{3}}}\int^{L/2}_{-L/2}\int^{L/2}_{-L/2}\int^{L/2}_{-L/2}
    \Psi^{2}(\textbf{r})\frac{\cos(q_{x}x)\cos(q_{y}y)\cos(q_{z}z)}{q}dxdydz.
\end{equation}
The first term in the hamiltonian (\ref{2.16}) describes harmonic
oscillations (optical phonons) about equilibrium positions
$Q_{\textbf{q}}=0$ (zero oscillations and excited phohons if they
are). The second term  $3\hbar^{2}/4m_{\ast}r_{0}^{2}$ is a
kinetic energy of an electron in the localized state. The last
term is a potential energy of an electron's interaction with a
deformation field. Following \cite{mahan} let us suppose the
equilibrium displacement is $\delta
Q_{\textbf{q}}=L_{\textbf{q}}/\hbar\omega$. Then the term in the
electron's energy which is linear in $Q_{\textbf{q}}$ disappears.
This gives us the energy of a system electron+deformation field
$I(r_{0})$:
\begin{eqnarray}\label{2.18}
H(Q_{\textbf{q}})=\frac{\hbar\omega}{2}\sum_{\textbf{q}}\left(|Q_{\textbf{q}}+\delta
Q_{\textbf{q}}|^{2}+|P_{\textbf{q}}|^{2}\right)
+\frac{3\hbar^{2}}{4mr_{0}^{2}}+\frac{1}{2\hbar\omega}\sum_{\textbf{q}}L_{\textbf{q}}^{2}\\
\equiv
\frac{\hbar\omega}{2}\sum_{\textbf{q}}\left(|Q_{\textbf{q}}+\delta
Q_{\textbf{q}}|^{2}+|P_{\textbf{q}}|^{2}\right)
+I(r_{0}).\label{2.18a}
\end{eqnarray}
The first term in (\ref{2.18}) describes the harmonic oscillations
about new equilibrium positions $-\delta Q_{\textbf{q}}$. The
second term is a kinetic energy of an electron in the localized
state. After the done transition in the hamiltonian (\ref{2.16})
the last term is a potential energy of an electron's interaction
with a deformation field plus an energy of the deformation field.

\begin{figure}[h]
\includegraphics[width=8.6cm]{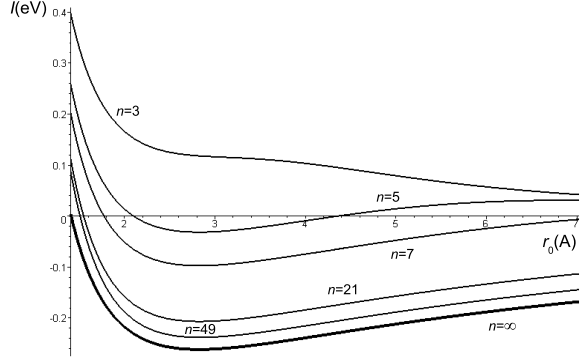}
\caption{The graphics of a dependence of a system's energy
electron+cluster (\ref{1.18}) on a polaron radius $I=I(r_{0})$ for
an ionic cluster consisting of $\texttt{NaCL}$. The curves were
build for edges of cubic clusters: $3a,5a,7a,21a,49a$, where
$a=2.81\texttt{A}$ is interatomic distances. The bold line is an
energy of an infinity crystal with an electron.} \label{fig7}
\end{figure}
\begin{figure}[h]
\includegraphics[width=8.6cm]{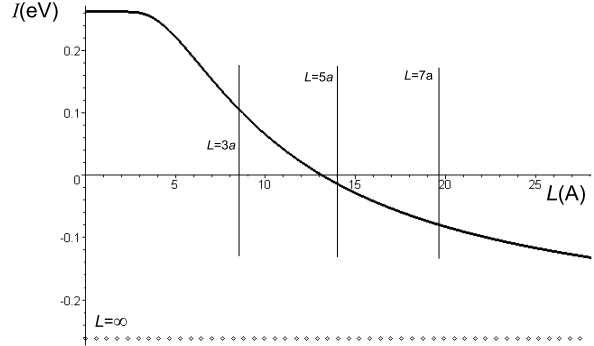}
\caption{The energy of a system electron+cluster (consisting of
$\texttt{NaCl}$) as a function of a diameter of the cluster $I(L)$
(solid line). The energy of a polaron in an infinity crystal
$L=\infty$ is shown by a dotted line. Lengthes of edges (in
interatomic distances $a$) of cubic clusters $n=3$, $n=5$, $n=7$
are marked by vertical lines.} \label{fig8}
\end{figure}

Summation over $\textbf{q}$ is equivalent to summation over
$n_{x},n_{y},n_{z}$ in accordance with Eq.(\ref{2.7}). Since the
summation is spreading on odd $n$ only, then it is convenient to
passing to new variables:
$\widetilde{n}_{x}=(n_{x}-1)/2,\widetilde{n}_{y}=(n_{y}-1)/2,\widetilde{n}_{z}=(n_{z}-1)/2$.
Besides in the formula (\ref{2.17}) the limits of integration can
be expanded to $\pm\infty$ that simplifies the calculation without
a essential error because the restriction (\ref{2.7}) on wave
vectors of phonons does a main contribution in the effect of a
finite volume of a cluster. Hence the energy can be written in a
view:
\begin{eqnarray}\label{2.19}
I(r_{0})=&&\frac{3\hbar^{2}}{4m_{\ast}r_{0}^{2}}+\frac{M_{0}^{2}}{2\hbar\omega}\frac{2}{L^{3}}
\left\{%
\begin{array}{cc}
  \sum_{\widetilde{n}_{x}=(-n+1)/2}^{(n-1)/2}\sum_{\widetilde{n}_{y}=(-n+1)/2}^{(n-1)/2}\sum_{\widetilde{n}_{z}=(-n+1)/2}^{(n-1)/2}
  & ,\texttt{for}\qquad \texttt{odd}\qquad n \\
  \\
  \sum_{\widetilde{n}_{x}=-n/2}^{(n-2)/2}\sum_{\widetilde{n}_{y}=-n/2}^{(n-2)/2}\sum_{\widetilde{n}_{z}=-n/2}^{(n-2)/2}
  & ,\texttt{for}\qquad \texttt{even}\qquad n \\
\end{array}%
\right\}\nonumber\\
&&\times\frac{L^{2}}{\pi^{2}}
\frac{\exp\left(-\frac{\pi^{2}}{L^{2}}\frac{r_{0}^{2}}{4}[(2\widetilde{n}_{x}+1)^{2}+(2\widetilde{n}_{y}+1)^{2}+(2\widetilde{n}_{z}+1)^{2}]\right)}
{(2\widetilde{n}_{x}+1)^{2}+(2\widetilde{n}_{y}+1)^{2}+(2\widetilde{n}_{z}+1)^{2}}
\end{eqnarray}
For an infinity crystal ($n\rightarrow\infty$) the system's energy and
the polaron radius are:
\begin{eqnarray}
I(r_{0})&=&\frac{3\hbar^{2}}{4mr_{0}^{2}}-\frac{M_{0}^{2}}{2\sqrt{2}\pi^{(3/2)}\hbar\omega}\frac{1}{r_{0}}\label{2.20}
\nonumber\\
r_{0}&=&\frac{3\sqrt{2}\pi^{(3/2)}\hbar^{2}\hbar\omega}{mM_{0}}.\label{2.21}
\end{eqnarray}
The results of calculation of $I(r_{0})$ for different $n$ are
shown in Fig.(\ref{fig7}). We can see for clusters with a length
of an edge $n>5$ ($n$ is number of interatomic distances
$a=2.81\texttt{A}$) a bound state of an electron exists, moreover
the polaron radius nearly does not depend on a size of a cluster
and it is equal to the polaron radius in an infinity crystal
(\ref{2.21}) $r_{0}\approx 2.8\texttt{A}$. With increase of a
cluster's size an energy of a polaron aspires to its energy in an
infinity crystal $I_{\infty}=-0.26\texttt{eV}$. Compared
Fig.(\ref{fig7}) with the results in Fig.(\ref{fig1}) obtained in
the quasiclassical approach we can see the quantization of
cluster's oscillation brings to the same critical diameter of a
cluster $5a$.

The bulky expression (\ref{2.19}) containing a triple sum can be
simplified by a continuous approximation. The polaron radius
$r_{0}$ nearly does not depend on a cluster's size $L$ and it is
equal to its value in an infinity crystal. Then in the expression
(\ref{2.19}) we can suppose $r_{0}$ equals to the optimal value
(\ref{2.21}). Hence the energy of a cluster with an electron is a
function of the cluster's size only: $I=I(L)$. In a cube cluster
with a length of an edge $L$ a phonon with a wave vector $q<\pi/L$
cannot propagate. The upper limit of wave vectors' values caused
by a crystal lattice does not influence on the result essentially.
Then the cluster can be supposed spherical with a diameter $L$ and
we can replace the triple sum (\ref{2.19}) by an integral:
\begin{equation}\label{2.22}
    I(L)=\frac{3\hbar^{2}}{4mr_{0}^{2}}+
    \frac{M_{0}^{2}}{2\hbar\omega\pi^{2}}\int_{\pi/L}^{\infty}\exp\left(-\frac{q^{2}r_{0}^{2}}{2}\right)dq.
\end{equation}
A result of the calculation of an energy of the system
electron+cluster with help of the formula (\ref{2.22}) is shown in
Fig.(\ref{fig8}). Let us compare the result with an exact result
shown in Fig.(\ref{fig7}). We can see the critical sizes of a
cluster in both methods of calculation coincide:
$L_{\texttt{cr}}=4a\div 5a$, energies of the system for
corresponding sizes of a cluster are nearly equal. At increase of
a cluster's size to infinity the energy of a polaron aspires to
its value in an infinity crystal. These facts afford ground to use
the simplified expression (\ref{2.22}) instead of the bulky
formula (\ref{2.19}).

\section{The quantum approach. A weak coupling theory of an electron in a cluster.}\label{weak}

Localization of an electron in a cluster (crystal) because of an
interaction with optical phonons can be described by a
perturbation theory. This supposes a mass operator or an effective
mass of an electron have peculiarities at the localization of the
electron. Further application of the perturbation theory for a
system with values of parameters (electron phonon coupling
constant, size of cluster) exceeding the critical values is
impossibly.

Let $G_{0}$ and $G$ are a free propagator and a dressed propagator
of an electron accordingly (here $\hbar=1$):
\begin{equation}\label{3.1}
    G_{0}(\textbf{k},\varepsilon)=\frac{1}{\varepsilon-\frac{k^{2}}{2m}+i\delta},\qquad
G(\textbf{k},\varepsilon)=\frac{1}{G_{0}^{-1}-\Sigma(\textbf{k},\varepsilon)},
\end{equation}
where $\delta\rightarrow 0^{+}$, $\Sigma(\textbf{k},\varepsilon)$
is a self-consistent mass operator:
\begin{equation}\label{3.2}
    -i\Sigma(\textbf{k},\varepsilon)=\int\frac{d\omega}{2\pi}\sum_{\textbf{q}}\frac{M_{0}^{2}}{Vq^{2}}
    iG(\textbf{k}-\textbf{q},\varepsilon-\omega)(-i)D_{0}(\textbf{q},\omega),
\end{equation}
$D_{0}(\textbf{q},\omega)$ is a propagator of a optical phonon:
\begin{equation}\label{3.3}
    D_{0}(\textbf{q},\omega)=\frac{2\omega_{0}}{\omega^{2}-\omega^{2}_{0}+2i\delta\omega_{0}}.
\end{equation}
We suppose optical oscillations are dispersionless
$\omega\neq\omega(q)$. After integration in (\ref{3.2}) over the
energetic parameter $\omega$ we have:
\begin{equation}\label{3.4}
\Sigma(\textbf{k},\varepsilon)=-\frac{M_{0}^{2}}{V}\sum_{\textbf{q}}\frac{1}{q^{2}}
\frac{1}{\omega_{0}+\frac{q^{2}}{2m}-\frac{\textbf{kq}}{m}-\left[\varepsilon-\frac{k^{2}}{2m}-\Sigma(\textbf{k})\right]}
\end{equation}
If in a crystal one electron propagates only then a
corresponding quasi-particle (dressed electron) coincides with the
particle (in a many-particle system quasi-particles are not
identical to particles of the system). This means the
quasi-particle has an infinity lifetime, hence it propagates on a
mass shell:
$\varepsilon=\frac{q^{2}}{2m}+\Sigma(\textbf{k},\varepsilon)$.
This equality brings to satisfaction to the self-consistence
conditions (\ref{3.4}).

\begin{figure}[h]
\includegraphics[width=8.6cm]{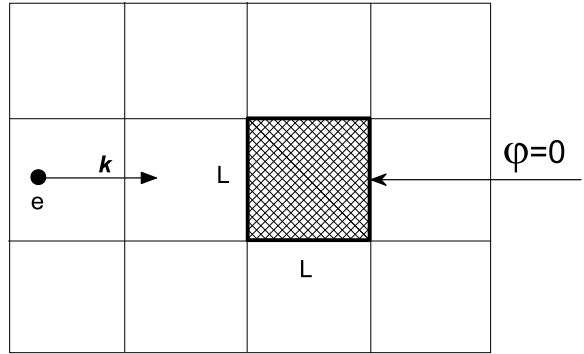}
\caption{A division of the crystal into cubes with an edge $L$. In
the system a charge $e$ is propagating with a momentum
$\textbf{k}$. On each cube face the boundary condition (\ref{2.5})
has been given - a potential of the polarization field
$\varphi=0$.} \label{fig9}
\end{figure}

For a crystal the summation over $\textbf{q}$ is done over all
vector space. As consequence of the boundary conditions
(\ref{2.5},\ref{2.5a}) on a surface of a cluster the summation
must be done over the wave vectors (\ref{2.7}) only. The system
must be homogeneous so that a notion of the mass operator $\Sigma$
has a sense. But one cluster in any medium is a spatially
inhomogeneous system. We propose a method shown in
Fig.(\ref{fig9}). Let us divide an ionic crystal into cubes with
an edge $L$ (a size of the cluster). Then let us introduce the
boundary conditions (\ref{2.5}) on each cube faces as on a
cluster's surface - a potential of the polarization field
$\varphi=0$. Let an electron propagates in this quasi-homogeneous
system and interacts with optical phonons. The boundary conditions
are imposed on the optical phonons. Then we can calculate the mass
operator because a diagram series can be uncoupled (a momentum is
conserved).

In a section (\ref{strong}) we showed a summation over the vectors
(\ref{2.7}) can be replaced by an integration over $q$ from
$\pi/L$ to $+\infty$. Then the mass operator has a view:
\begin{eqnarray}\label{3.5}
  \Sigma &=& \frac{M_{0}}{(2\pi)^{3}}\int^{-1}_{1} dx\int_{\pi/L}^{\infty}\frac{dq}{\frac{q^{2}}{2m}-\frac{kq}{m}x+\omega_{0}} \nonumber\\
  &=& -\frac{2\pi m}{k}\arcsin\frac{k}{\sqrt{2m\omega_{0}}}-\int^{-1}_{1}\frac{2mdx}{\sqrt{2m\omega_{0}-k^{2}x^{2}}}
  \arctan\left(\frac{\pi/L-kx}{\sqrt{2m\omega_{0}-k^{2}x^{2}}}\right).
\end{eqnarray}
The second term determines the finite size correction to the mass
operator $ \Sigma_{\infty}=-\frac{2\pi
m}{k}\arcsin\frac{k}{\sqrt{2m\omega_{0}}}$. Supposing
$L\rightarrow\infty$ we have $\Sigma=\Sigma_{\infty}$.

\begin{figure}[h]
\includegraphics[width=8.6cm]{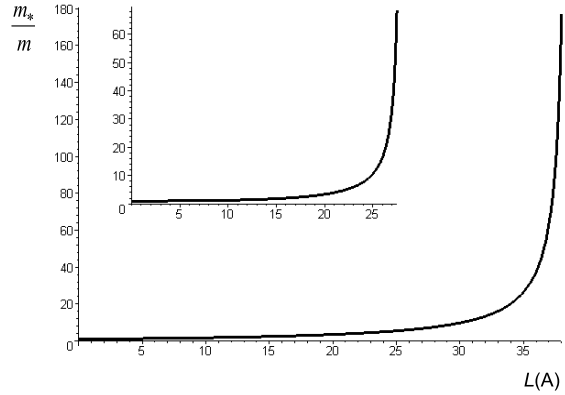}
\caption{The dependence of an electron's effective mass on a size
of a cluster (consisting of $\texttt{NaCl}$) $m_{\ast}(L)/m$ in a
limit $k\rightarrow 0$. The critical cluster's size is
$L_{\texttt{cr}}\approx 39\texttt{A}\approx 14a$. A graph on the
insert is a dependence $m_{\ast}(L)/m$ caused by the main
contribution
$\frac{2}{\pi}\arctan\frac{\pi/L}{\sqrt{2m\omega_{0}}}$ only. In
this case the critical cluster's size is $L_{\texttt{cr}}\approx
28\texttt{A}\approx 10a$. }\label{fig10}
\end{figure}

A dispersion law of a quasiparticle is determined by a pole
of the propagator $G$: $\varepsilon=\frac{k^{2}}{2m}+\Sigma(k)$.
For the limit $k\rightarrow 0$ we have an expression:
\begin{eqnarray}\label{3.6}
  \varepsilon=&&-\alpha\omega_{0}\left(1-\frac{2}{\pi}\arctan\frac{\pi/L}{\sqrt{2m\omega_{0}}}\right)\nonumber\\
  &&+\frac{k^{2}}{2m}\left[1-\frac{\alpha}{6}\left(1-\frac{2}{\pi}\arctan\frac{\pi/L}{\sqrt{2m\omega_{0}}}-
  \frac{2\sqrt{2m\omega_{0}}}{L}\frac{\pi^{2}/L^{2}-2m\omega_{0}}{(\pi^{2}/L^{2}+2m\omega_{0})^{2}}\right)\right]\nonumber\\
  &&\equiv-\alpha\omega_{0}\left(1-\frac{2}{\pi}\arctan\frac{\pi/L}{\sqrt{2m\omega_{0}}}\right)+\frac{k^{2}}{2m_{\ast}},
\end{eqnarray}
where we introduced an effective mass of an electron $m_{\ast}$:
\begin{equation}\label{3.7}
    m_{\ast}=\frac{m}{1-\frac{\alpha}{6}\left(1-\frac{2}{\pi}\arctan\frac{\pi/L}{\sqrt{2m\omega_{0}}}-
  \frac{2\sqrt{2m\omega_{0}}}{L}\frac{\pi^{2}/L^{2}-2m\omega_{0}}{(\pi^{2}/L^{2}+2m\omega_{0})^{2}}\right)}.
\end{equation}
From (\ref{3.6}) we can see a bottom of a conduction band falls.
At $L\rightarrow\infty$ a fall of the bottom correspond to a
result for an infinity crystal $-\alpha\omega$. The expression for
an effective mass (\ref{3.7}) is more interesting. In an infinity
crystal we have the expression
$m_{\ast}=m/\left(1-\frac{\alpha}{6}\right)$. At $\alpha=6$ the
effective mass turns into an infinity $m_{\ast}(\alpha=6)=\infty$,
that means the autolocalization of an electron. At $\alpha>6$ the
perturbation theory is not applicable. If the crystal is finite
(cluster as in Fig.\ref{fig9}) then the effective mass is a
function of a cluster's size $m_{\ast}(L)$. In an above-mentioned
domain of values of a electron-phonon coupling constant such a
value $L_{\texttt{cr}}$ exists that
$m_{\ast}(L_{\texttt{cr}})=\infty$,
$m<m_{\ast}(L<L_{\texttt{cr}})<\infty$, $m_{\ast}(0)=m$. Thus if a
cluster consists of a substance with $\alpha>6$ then a critical
size of the cluster exists when an electron can be autolocalized.
A dependence of a electron's effective mass on a size of a cluster
consisting of $\texttt{NaCl}$ is shown in Fig.(\ref{fig10}). The
term $\frac{2}{\pi}\arctan\frac{\pi/L}{\sqrt{2m\omega_{0}}}$ gives
a main contribution in the dependence (\ref{3.7}) because
\begin{eqnarray}\label{3.7b}
&&\lim_{L\rightarrow
0}\frac{2}{\pi}\arctan\frac{\pi/L}{\sqrt{2m\omega_{0}}}=1,\qquad
\lim_{L\rightarrow
\infty}\frac{2}{\pi}\arctan\frac{\pi/L}{\sqrt{2m\omega_{0}}}=0
\nonumber\\
&&\lim_{L\rightarrow
0}\frac{2\sqrt{2m\omega_{0}}}{L}\frac{\pi^{2}/L^{2}-2m\omega_{0}}{(\pi^{2}/L^{2}+2m\omega_{0})^{2}}=0,\qquad
\lim_{L\rightarrow
\infty}\frac{2\sqrt{2m\omega_{0}}}{L}\frac{\pi^{2}/L^{2}-2m\omega_{0}}{(\pi^{2}/L^{2}+2m\omega_{0})^{2}}=0
  \nonumber,
\end{eqnarray}
that is just this term gives a required asymptotical dependence on
$L$:
\begin{eqnarray}\label{3.7a}
    &&\Sigma(L\rightarrow\infty)=\Sigma_{\infty},\qquad \Sigma(L\rightarrow
    0)=0\nonumber\\
&&m_{\ast}(L\rightarrow\infty)=m_{\ast}(\infty),\qquad
m_{\ast}(L\rightarrow 0)=m
\end{eqnarray}
If to take into account this contribution only then we will obtain
the same picture qualitatively with a not large divergence of the
critical size.

\begin{figure}[h]
\includegraphics[width=8.6cm]{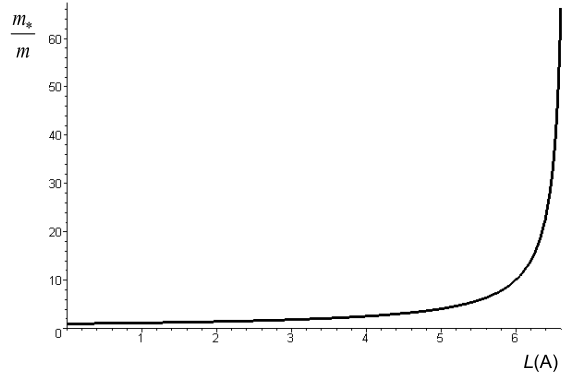}
\caption{The dependence of electron's effective mass on a size of
the cluster $m_{\ast}(L)/m$ in a limit $k\rightarrow
\sqrt{2m\omega_{0}}$. The critical cluster's size is
$L_{\texttt{cr}}\approx 6.8\texttt{A}\approx 2a\div
3a$.}\label{fig11}
\end{figure}

Above we considered a limit $k\rightarrow 0$. But the further
consideration shows the effective mass is a function of an
electron's momentum. In a domain $k^{2}/2m<\omega_{0}$ for an
infinity crystal we have:
\begin{eqnarray}\label{3.8}
\varepsilon_{\infty}=&&\frac{k^{2}}{2m}+\Sigma_{\infty}(k)=-\alpha\omega_{0}+\frac{k^{2}}{2m}
\left(1-\frac{\alpha}{6}-\frac{3\alpha}{40}\frac{k^{2}}{2m\omega_{0}}-\frac{15\alpha}{336}\frac{k^{4}}{(2m\omega_{0})^{2}}-...\right)\nonumber\\
&&\Rightarrow
m_{\ast}=\frac{m}{1-\alpha\frac{2m\omega_{0}}{k^{2}}\left[\frac{\sqrt{2m\omega_{0}}}{k}\arcsin\frac{k}{\sqrt{2m\omega_{0}}}-1\right]}
\Rightarrow
m_{\ast}(k\rightarrow\sqrt{2m\omega_{0}})=\frac{m}{1-\alpha\left(\frac{\pi}{2}-1\right)}.
\end{eqnarray}
In a domain $k^{2}/2m>\omega_{0}$ Cherenkov radiation of a phonon
is possible. Let us generalize the expressions for an finite
crystal with help of the expression (\ref{3.5}). A main
contribution into the dependence on cluster's size $L$ (a length
of a cube edge or a diameter) is given by the expression
$\frac{2}{\pi}\arctan\frac{\pi/L}{\sqrt{2m\omega_{0}}}$. Then we
obtain:
\begin{eqnarray}\label{3.9}
&&\Sigma(L)=\Sigma\infty\left(1-\frac{2}{\pi}\arctan\frac{\pi/L}{\sqrt{2m\omega_{0}}}\right),\nonumber\\
&&m_{\ast}(k\rightarrow
0)=\frac{m}{1-\frac{\alpha}{6}\left(1+\frac{9}{20}\frac{k^{2}}{2m\omega_{0}}+\frac{15}{56}\frac{k^{4}}{(2m\omega_{0})^{1}}+\ldots\right)
\left(1-\frac{2}{\pi}\arctan\frac{\pi/L}{\sqrt{2m\omega_{0}}}\right)},\nonumber\\
&&m_{\ast}(k\rightarrow\sqrt{2m\omega_{0}})=\frac{m}{1-\alpha\left(\frac{\pi}{2}-1\right)
\left(1-\frac{2}{\pi}\arctan\frac{\pi/L}{\sqrt{2m\omega_{0}}}\right)}
\end{eqnarray}
A dependence of an electron's effective mass on a cluster's size
in a limit $k\rightarrow\sqrt{2m\omega_{0}}$ is shown in
Fig.(\ref{fig11}). At these momentums of an electron a more strong
coupling with optical phonons takes place. Hence the critical
diameter of a cluster is  much less than one in the limit
$k\rightarrow 0$. Thus the weak coupling approach gives a
dependence of the critical size on an electron's momentum
$L_{\texttt{cr}}(k)$. However the critical sizes in this
approaches are several interatomic distances ($\propto 5a\div
10a$) as in the strong coupling approach.

\section{Conclusion.}\label{concl}

In this article we calculated a polaron's energy in an ionic
cluster (by the example of a nanoparticle consisting of
$\texttt{NaCl}$) and a critical size of the cluster regarding in
formation of a autolocalized state of an additional electron. The
calculation is done by both the quasiclassical method and the
quantum approach (in a sense of quantization of a deformation
field). Moreover we generalized a weak coupling method for a case
of a polaron in a cluster. The results of the work are following.

As in numerically solution and computer simulation in
\cite{lakhno} we obtained that the bound state of an electron in
clusters from polar substances can be realized starting with a
some critical radius in consequence with the boundary conditions
on a cluster's surface for a potential of the polarization field
$\varphi(R)=0$. With increase of a cluster's size the binding
energy increases and aspires to a polaron energy in an infinity
crystal. A radius of an electron's autolocalization in a cluster
(a polaron size) is such that an energy of a system
electron+polarized cluster is minimal. However by the analytical
variational method we found that this radius nearly does not
depend on the cluster's radius. The binding energy increases
monotonically with increase of the cluster's size. Moreover the
pattern is observed: the bound state exists while the polaron
radius is less than the cluster's radius. That is an autolocalized
electron is hidden in a depth of the cluster and has a negligibly
low probability to be on the outside. If the cluster is enough
small so that an electron can be on the outside the bound state
disappears. Unlike results of numerical simulation in
\cite{lakhno,bar1,bar2} where surface states were found which can
be understood as excited state we found that if the polarization
well is enough deep (cluster's size is enough big) then except
$s$-state of the autolocalized electron other discrete levels are
possible. We calculated energy of $p$-state and obtained a
cluster's critical radius for this state. It is necessary to
notice that similar results ware obtained in \cite{bar0}, but
another functional $I(\Psi,\varphi)$ was considered and ground
state energy of an electron in a cluster and the autolocalization
radius were expressed by static dielectric constant only. Hence in
this model autolocalisation effect is not polaron effect unlike
our model.

An interaction of an electron with the polarization field is the
interaction with longitudinal optical phonons. Due an ionic type
connection between atoms ionic clusters form stable cuboid
structures and have a cubic microlattice. However it is necessary
to notice for clusters with a size bigger the the critical size
$R\gg r_{0}$ an excess electron has low probability to be near a
surface of the cluster hence the shape have not principal
importance. The boundary condition on a cluster's surface for the
polarization field $\varphi(R)=0$ brings to a restriction on
possible wave vectors of optical phonons $\pi/L<q<\infty$. The
optical phonons are standing waves between opposite sides of a
nanoparticle. We obtained the polaron energy in a cluster
(\ref{2.19}) taking into account the restrictions on a propagation
of phonons. Results for an energy and a radius of the polaron
correspond to the results of the quasiclussical approach -
existence of a critical size of a cluster regarding in formation
of an autolocalized state of an electron, an asymptotical
aspiration of a binding energy of an electron in a cluster to the
binding energy in an infinity crystal with an increase of a
cluster's radius, an independent of a polaron radius on a
cluster's size, an infinity low probability for an electron to be
on the outside of a cluster if the electron is in the
autolocalized state. These results are different from the results
\citep{land,honea1,rajag} of computer simulation of electron
localization in alkali-halide clusters $M_{n}^{+}X_{n-1}^{-}$ and
$M_{n}^{+}X_{n-2}^{-}$ which contain F-center defects and an
electron localized near halide ion. Unlike our result, where an
excess electron autolocalizes on a neutral ionic cluster due
interaction with the induced polarization, such systems have not
any critical parameters. This problem is analogous to a problem of
F-center model of negatively charged metal-ammonium clusters
\cite{lakhno2}. It is necessary to notice that an essential
difference of a polaron state in a cluster from the polaron state
in a crystal is localization of an electron in the center of a
cluster. On the contrary in an infinity crystal all points are
equivalent (in a continual representation of a medium) hence the
electron can move.

We described an electron's localization in a cluster due an
interaction with longitudinal optical phonons by a perturbation
method taking into account the boundary conditions for the
electric field of a polarization. The electron localization in a
cluster manifests itself as a singularity of an electron's
effective mass $m(L_{\texttt{cr}})=\infty$. Using the perturbation
method we showed a dependence of the critical size of a cluster on
a electron's momentum in a conduction band
$L_{\texttt{cr}}=L_{\texttt{cr}}(k)$. The largest critical size of
a cluster is for an electron situated on a bottom of the
conduction band, the smallest critical size is for an electron's
momentum on a limit of Cherenkov radiation of an optical phonon.
This fact is explained by the fact that an interaction of an
electron with an ionic crystal increases with an increase of an
electron's kinetic energy, because the larger electron's energy
the stronger an electron radiates virtual phonons and then
interacts with them.

\end{document}